\documentclass[prd,aps,showpacs,showkeys]{revtex4-1}
\usepackage{bm,amsfonts,latexsym,amsmath,amssymb,amsbsy,graphicx}

\newcommand{\bb}{\begin{eqnarray}}
\newcommand{\ee}{\end{eqnarray}}
\newcommand{\ba}{\begin{align}}
\newcommand{\ea}{\end{align}}

\begin{document}

\title{\bf  Creation of planar charged fermions in Coulomb and Aharonov-Bohm potentials}
\author{V.R. Khalilov}\email{khalilov@phys.msu.ru}
\affiliation{Faculty of Physics, Moscow State University, 119991,
Moscow, Russia}

\begin{abstract}
The creation of charged fermions from the vacuum by a Coulomb field
in the presence of an Aharonov--Bohm (AB)
potential are studied  in  2+1 dimensions. The process is governed
by a (singular) Dirac Hamiltonian that requires
the supplementary definition in order for it to be treated
as a self-adjoint quantum-mechanical operator.
By constructing a one-parameter self-adjoint extension of
the Dirac Hamiltonian,  specified by boundary conditions,
we describe the (virtual bound) quasistationary  states
with ``complex energy'' emerging in an attractive Coulomb potential,
derive for the first time,  complex equations (depending upon the electron spin
 and the extension parameter) for the quasistationary  state
``complex energy''. The constructed self-adjoint Dirac Hamiltonians in
Coulomb and  AB potentials are applied to provide a correct
description to the low-energy electron excitations,
as well as  the creation of charged quasiparticles from the vacuum in graphene
by the Coulomb impurity in the presence of AB potential.
It is shown that the strong Coulomb field can create charged fermions  for some range
of the extension parameter.
\end{abstract}



\maketitle

\section{Introduction}

The creation of electron--positron  pairs from the vacuum
by a Coulomb  field is an important effect of quantum electrodynamics
 exhaustively studied  in \cite{wpwg,sgyz,zelpop,abm,jrlfak,msbmwg,blp,grrein}.
We recall that the energy spectrum of an electron in the Coulomb four-potential $A^0(r)=a/e_0r, {\bf A}=0, a>0, e_0>0$, ($-e_0$ is the electron charge) with $1>a$ consists of a continuous spectrum $|E|\geq m$ separated by a gap $2m$ ($m$ is the electron mass, and we use the system of units with $c=\hbar=1$) and a discrete spectrum $0 < E < m$ inside the gap. The formula for the lowest energy level $E=m\sqrt{1-a^2}$ formally gives imaginary eigenvalues for the Dirac Hamiltonian with $a>1$.
As for the Dirac equation with $a>1$,  it was considered inconsistent and physically
meaningless. The difficulty of the imaginary spectrum in the case of $a>1$ are solved (see, \cite{grrein}) by replacing the singular potential $a/e_0r$
 with a  Coulomb potential cut off at short distance $R$ for which the Dirac equation has physically meaningful solutions. In a (cut off) Coulomb potential, as $a$ increases,
 the lowest electron energy level becomes negative for
$a > 1$ descends to the upper boundary $E=-m$ of the
lower continuum at $a=a_{cr}$, and can dive into it for $a>a_{cr}$, signaling  the instability
of the quantum electrodynamic vacuum in the overcritical Coulomb field.
 The lowest state then turn into resonance with a finite lifetime which
can be described as a quasistationary state with ``complex energy''. The so-called critical  charge $a_{cr}$ is defined as the condition for the appearance of the imaginary part of ``the energy''.
The latter is related to the total probability of the creation of  electron-–positron pairs by the overcritical Coulomb field:  the positron goes to infinity and the electron is coupled to the Coulomb center. Thus, the problem can no longer be considered a one-particle one.

In the problem on a massive charged fermion in a strong (cut off) Coulomb field in
2+1 dimensions, the picture is similar, but the ground-state energy vanishes at $a=1/2$ \cite{khho,kh2009}. The case of massless charged fermions also is of great interest.
Close to the so-called Dirac points,
charged quasiparticle excitations in the potential of graphene lattice
are massless Dirac-like fermions characterized by a linear dispersion
relation \cite{netall,ngpng,kupgc} and so a single electron dynamics in graphene
is described by a massless two-component Dirac equation \cite{ngpng,ksn,zji,vpnn,ashkl,ifh}.
This allows to consider graphene as the condensed matter analog for
relativistic quantum field theory \cite{datdc} and
massless charged  quasiparticles in graphene \cite{gnn} can provide an
interesting realization of quantum electrodynamics
in 2+1 dimensions \cite{ggvo,ggvo1}.
Since, the ``effective fine structure constant" in graphene  is large,
 there appears a new possibility to study a strong-coupling version of the quantum
electrodynamics (QED) and the existence of charged Fermi quasiparticles
in graphene makes  experimentally feasible to observe the creation of quasiparticles
by  static electric fields \cite{datdc}.

For massless fermions, there are no discrete levels in the cut off Coulomb potential
due to scale invariance of the massless Dirac equation,
nevertheless for $a>1$ quasistationary states emerge \cite{kupgc,ashkl,skl2,ggg,gs}.
It should be noted that the induced current in
graphene in the field of solenoid was found to be a
finite periodical function of the magnetic flux \cite{jmpt} and
Coulomb impurity problems, such as the vacuum polarization and screening, in graphene were 
studied in \cite{vpnn,ashkl,tmksh}. The creation of graphene quasiparticles
from vacuum by the space homogeneous static electric field was
studied in \cite{datdc} by means of the methods of planar quantum electrodynamics
developed in \cite{sgdm,glc,cbpges,ggbpg}.

The above-mentioned difficulties do not arise if the Dirac
Hamiltonian with the no cutoff Coulomb field
(and with arbitrary $a$) is correctly defined
as a self-adjoint operator.
By constructing of the self-adjoint Dirac Hamiltonians
using the so-called form asymmetry
method developed in  \cite{vgt,gtv1}, here
 we investigate the creation of  charged fermions
from the vacuum  by a Coulomb field in the presence of AB
potential in 2+1 dimensions. We show that there exists a family
of self-adjoint Dirac Hamiltonians parameterized by an
extension parameter (and specified by boundary conditions
at the singular point) and a set of  quasistationary states with ``complex energies''  can
be evaluated for each Hamiltonian (see, also \cite{gtv1}). The different boundary
conditions on the  wave functions imposed at the origin are of importance leading
to inequivalent physical cases in the relevant two spatial dimensions.
The presence of AB potential  allows us to study
the influence of the particle spin on the physical effects, which is due
to the interaction between the electron spin magnetic moment
and the Aharonov-Bohm magnetic field \cite{KhaHo07}.
It  will be noted that the self-adjoint Dirac Hamiltonians in
2+1 dimensions   were constructed in \cite{phg,ampw,vrkh} for the AB problem, analyzed
in the nonrelativistic limit in \cite{safb} for the so-called Aharonov--Casher problem \cite{ahc} of the motion of a neutral fermion with an anomalous magnetic moment
in the electric field of an electrically charged conducting
long straight thin thread oriented perpendicularly to the plane of fermion motion;
particle creation in a moving cosmic string, governed by the Dirac Hamiltonian with AB potential
in 2+1 dimensions, was discussed in \cite{aw} The problems of self-adjointness
of the Dirac Hamiltonians with Aharonov-–Bohm and magnetic-solenoid fields were studied in \cite{sgdgas,ggsv0}.

\section{Solutions and spectra of the radial Dirac Hamiltonian. Self-adjoint boundary conditions}

The space of particle quantum states in two spatial dimensions  is
the Hilbert space $\mathfrak H=L^2(\mathbb R^2)$ of square-integrable
functions $\Psi({\bf r}), {\bf r}=(x,y)$ with the scalar product
\bb
(\Psi_1,\Psi_2)=\int \Psi_1^{\dagger}({\bf r})\Psi_2({\bf r})d{\bf r},\quad d{\bf r}=dxdy.
\label{scpr}
\ee
The Dirac equation for a fermion in a given external field can be obtained just as in 3+1 dimensions.

 First, we consider the massive case. The Dirac $\gamma^{\mu}$-matrix algebra
is known to be represented  in terms of the
two-dimensional Pauli matrices $\sigma_j$ and the parameter $s=\pm 1$
can be introduced to label two types of fermions
in accordance with the signature of the two-dimensional
Dirac matrices \cite{hoso} and  is applied to characterize two states
of the fermion spin (spin ``up" and ``down") \cite{crh,khlee}.
Then, the Dirac Hamiltonian
for a fermion of the mass $m$ and charge
$e=-e_0<0$ in an   Aharonov--Bohm
$A_0=0$, $A_r=0$, $A_{\varphi}=B/r$, $r=\sqrt{x^2+y^2}$, $\varphi=\arctan(y/x)$
and Coulomb $A_0(r) =a/e_0r$, $A_r=0$, $A_{\varphi}=0$, $a>0$
potentials, is
\bb
 H_D=\sigma_1P_2-s\sigma_2P_1+\sigma_3 m-e_0A_0(r),\label{diham}
\ee
where $P_\mu = -i\partial_{\mu} - eA_{\mu}$ is the
generalized fermion momentum operator (a three-vector).
The Hamiltonian (\ref{diham}) should
be defined as a self-adjoint operator in the Hilbert space
of square-integrable two-spinors $\Psi({\bf r}), {\bf r}=(x,y)$
with the scalar product (\ref{scpr}).
The total angular momentum $J\equiv L_z+ s\sigma_3/2$, where $L_z\equiv
-i\partial/\partial\varphi$, commutes with $H_D$, therefore, we can consider
separately in each eigenspace  of the operator $J$
 and the total Hilbert space is a direct orthogonal sum of subspaces of $J$.

To avoid misunderstanding, we note that by Coulomb potential in 2+1 dimensions, we mean potential
that decrease as $1/r$ with the distance from the source, having in mind that in a physical situation (e.g., in graphene), although the electrons move in a plane, their Coulomb interaction with the external field of the pointlike charge of an impurity occurs in a physical (three-dimensional) space and the electric field strength of the impurity is a three-dimensional (not two-dimensional) vector. Therefore, the
potential $A_0(r) \sim 1/r$ (and not $A_0(r) \sim \log r$, as would be
the case in 2+1 dimensions) does not satisfy the
two-dimensional Poisson equation with a pointlike source at the origin. Similarly, in real physical space, the quantity $B$ characterizes the flux  of the Aharonov--Bohm magnetic field
 ${\bf H}=(0,\,0,\,H)=\nabla\times {\bf A}= \pi B\delta({\bf r})$ and
 leads to the interaction potential of the electron spin magnetic moment with the
magnetic field in the form $-s eB \delta(r)/r$, which is singular and
   must influence the behavior of solutions at the origin.
The ``spin'' potential is invariant under the changes  $e\to -e, s\to -s$, and it hence suffices to consider only the case $e=-e_0<0$ and $eB\equiv -\mu<0$. Then, the potential is attractive for $s=-1$ and repulsive for $s=1$.

Eigenfunctions of the Hamiltonian (\ref{diham}) are (see, \cite{hkh,khlee1})
\bb
 \Psi(t,{\bf r}) = \frac{1}{\sqrt{2\pi r}}
\left( \begin{array}{c}
f_1(r)\\
f_2(r)e^{is\varphi}
\end{array}\right)\exp(-iEt+il\varphi)~, \label{three}
\ee
where $E$ is  the fermion energy, $l$ is an integer.
The wave function $\Psi$ is an eigenfunction of the
operator $J$ with eigenvalue $j=l+s/2$ and
\bb \check h F= EF, \quad F=\left(
\begin{array}{c}
f_1(r)\\
f_2(r)\end{array}\right), \label{radh}\ee
where
\bb
\check h=is\sigma_2\frac{d}{dr}+\sigma_1\frac{l+\mu+s/2}{r}+\sigma_3m-\frac{a}{r},\quad \mu\equiv e_0B
\label{radh0}
\ee
Thus, the problem is reduced to that for the radial Hamiltonian $\check h$
 in the Hilbert space of  doublets $F(r)$ square-integrable on the half-line.

As was shown in \cite{khlee,khlee1} any correct doublets $F(r)$, $G(r)$ of the Hilbert space
$\mathfrak H=\mathfrak L^2(0,\infty)$ must satisfy
\bb
\lim_{r\to 0} G^{\dagger}(r)i\sigma_2 F(r)=0. \label{bounsym}
\ee
Then, the needed solution of (\ref{radh}) is
\begin{align}
F&=e^{-x/2}x^{\gamma_s}\left[v^+\Phi(a^s,\;c_s\;;x)+v^-  m^+_s\Phi(a^s+s,\;c_s\;;x)\right]
\equiv Y(r,\gamma_s,E), \quad x=2\lambda r,\lambda = \sqrt{m^2-E^2}. \label{gensol}
\end{align}
Here $\gamma_s=\pm\sqrt{(l+\mu+s/2)^2-a^2}\equiv\gamma_s^{\pm}$,
$a^s=\gamma_s+\frac12-\frac{s}{2}-\frac{aE}{\lambda},\quad c_s=2\gamma_s+1$,
$m_s^{\pm}=[s\gamma_s^{\pm}-Ea/\lambda]/[\nu+am/\lambda]$,
\bb
v^+=
\left(\begin{array}{c}
	1 \\ p
\end{array}\right),\quad
v^-=
\left(\begin{array}{c}
	1 \\ -p
\end{array}\right),\quad
p=\sqrt{\frac{m-E}{m+E}},
\ee
$\Phi(a, c; x)$ is
the confluent hypergeometric function \cite{GR}.

We denote $\gamma_s^+=\sqrt{\nu^2-a^2}\equiv \gamma$ for $a^2\leq \nu^2$,    $\gamma_s^+=i\sqrt{a^2-\nu^2}\equiv i\sigma$ for $a^2>\nu^2$ and we note that
\bb
\gamma(\pm l, s=1, \mu, a)=\gamma(\pm l+1, s=-1, \mu, a).
\label{grel}
\ee
Then, for $\gamma\ne n/2$, $n=1, 2 ,\ldots$, the needed  linear independent solutions are:
\begin{align}
U_1(r;E)&=Y(r,\gamma_s,E)|_{\gamma_s=\gamma},
\nonumber \\
U_2(r;E)&=Y(r,\gamma_s,E)|_{\gamma_s=-\gamma}
\label{1e35}
\end{align}
with the asymptotic behavior at $r\to 0$
\begin{align}
U_1(r;E)&=r^{\gamma}u_+{+}O(r^{\gamma+1}),
\nonumber \\
 U_2(r;E)&=r^{-\gamma}u_-{+}O(r^{-\gamma+1}),
\label{e36}
\end{align}
where
\bb
u_{\pm}=\left(\begin{array}{c}
\displaystyle(\pm s\gamma+\nu)/a \\ 1
\end{array}\right),
\nonumber
\ee
as well as
\bb
V_1(r;E)=U_1(r;E)+\frac{a}{2s\gamma}\omega(E)U_2(r;E),
\label{e40}
\ee
where $\omega(E)={\rm Wr}(U_1,V_1)$ is the Wronskian:
\bb
\omega(E)=\frac{\Gamma(2\gamma)\Gamma\left(-\gamma+(1-s)/2-aE/\lambda\right)}{\Gamma(-2\gamma)
\Gamma\left(\gamma+(1-s)/2-aE/\lambda\right)}\frac{(2\lambda)^{-2\gamma}}{m^{-2\gamma}}
\frac{(1-m^-_s)}{(1-m^+_s)}\frac{2s\gamma}{a}\equiv\frac{\tilde{w}(E)}{\Gamma(-2\gamma)}.
\label{wren}
\ee

Any doublet of the domain $D(h)$ must satisfy
\bb
 (F^{\dagger}(r)i\sigma_2 F(r))|_{r=0}= (\bar f_1f_2-\bar f_2f_1)|_{r=0} =0. \label{bounsym1}
 \ee
The quantities $q=\sqrt{\nu^2-\gamma^2}$ and  $q_c=\nu \Leftrightarrow\gamma=0$ are called the effective and   critical charge, respectively; it is helpful also to determine $q_u=\sqrt{\nu^2-1/4}\Leftrightarrow\gamma=1/2$.
As was shown in \cite{khlee1} for $q\leq q_u$, $\gamma\geq 1/2$,  the domain
$D(h)$ is the space of absolutely continuous doublets $F(r)$  regular at $r=0$ with $hF(r)$ belonging to $\mathfrak L^2(0,\infty)$.

For $0<\gamma<1/2$ ($q_u<q<q_c$) there is one-parameter $U(1)$-family of the operators $h_{\theta}\equiv h_{\xi}$, $\xi=\tan(\theta/2), -\infty\leq\xi\leq+\infty, -\infty\thicksim{+\infty}$, with the domain  $D_{\xi}$
\bb
h_\xi{:}\left\{\begin{array}{l}
 D_\xi=\left\{\begin{array}{l}
F(r):F(r)\;\mbox{is absolutely continuous in}[0,\infty),
\nonumber \\
F,\check hF\in{\mathfrak L}^2(0,\infty), \\
F(r)=c[r^\gamma{u_+}-\xi r^{-\gamma}u_{-}]+O(r^{1/2}),\;|\xi|<\infty,\\
F(r)=cr^{-\gamma}u_-{+}O(r^{1/2}),\;r\rightarrow{0},\;\xi=\infty,
\end{array}\right.\\
h_\xi F=\check hF,
\end{array}\right.
\ee
where $c$ is arbitrary constant. The  operator
$h^0$ is not determined as an unique self-adjoint operator and so the additional
specification of its domain, given with the real parameter $\xi$ (the self-adjoint extension parameter)  is required in terms of the self-adjoint boundary conditions. Physically, the self-adjoint boundary conditions  show that the probability current density  is equal to zero at the origin.
\bb
\frac{d\sigma(E)}{dE}=\frac{1}{\pi}\lim\limits_{\epsilon\rightarrow{0}}{\rm Im}\frac{1}{\omega_{\xi}(E+i\epsilon)},
\label{specfun}\ee
where the generalized function $\omega_{\xi}(E+i\epsilon)$ is obtained by the analytic continuation of the corresponding Wronskian in the complex plane of $E$; on the real axis of $E$ it is just the function $\omega(E)$ determined by (\ref{wren}) for $\xi=0$. For $0<\gamma<1/2$ the doublet  $U_\xi(r;E)=U_1(r;E)-\xi U_2(r;E)$ and at $r\to 0$
 $U_\xi(r;E)=r^{\gamma}u_+-\xi r^{-\gamma}u_{-}+O(r^{-\gamma+1})$.
Solution $V_1$ is now  $V_1(r;E)\equiv V_\xi=U_\xi(r;E)+[a/2s\gamma]\omega_\xi(E)U_2(r;E)$
with $\omega_\xi(E)={\rm Wr}(U_\xi,V_\xi)=\omega(E)+2s\gamma\xi/a$
and $\omega(E)$ determined by (\ref{wren}). So  $\omega_{\xi}(E)=\lim\limits_{\epsilon\rightarrow{0}}{\omega}_{\xi}(E+i\epsilon)$
and, thus, the spectral function is determined by the generalized function $F(E)=\lim\limits_{\epsilon\rightarrow{0}}{\omega}_{\xi}^{-1}(E+i\epsilon)$.
At the points, at which the function
${\omega}_{\xi}(E)=\lim\limits_{\epsilon\rightarrow{0}}{\omega}_{\xi}(E+i\epsilon)$
is not equal zero $F(E)=1/\omega_{\xi}(E)$.  It can be verified that in the range $|E|>m$
the functions $\omega(E)$  and $\omega_{\xi}(E)$ are
continuous, complex-valued and  not equal to zero for real $E$;
the spectral function $\sigma(E)$ exists and is absolutely continuous.
Thus, the energy spectrum in the range $|E|\geq m$ is continuous.
In the range $|E|<m (-m<E<m)$ the functions $\omega(E)$  and $\omega_{\xi}(E)$ are real and $\lim\limits_{\epsilon\rightarrow{0}}{\omega}_{\xi}^{-1}(E+i\epsilon)$ can be complex only at the points where $\omega_{\xi}(E)=0$ and the energy spectrum  of bound states is determined by roots of this equation. The Wronskians as a function of the complex $E$ involve  $\lambda=\sqrt{m^2-E^2}$, have two cuts $(-\infty, -m]$ and $[m,\infty)$
in the complex plane of $E$ and two sheets: ${\rm Re}\lambda>0$, the first (physical) sheet and ${\rm Re}\lambda<0$, the second (unphysical) sheet). Bound states are situated on the physical sheet of $\lambda$. For $\gamma\geq 1/2$ the discrete spectrum is
\cite{khlee1}
 \bb
 E_{n,l} = m\frac{n + (1-s)/2 + \sqrt{\nu^2-a^2}}
 {\sqrt{[n + (1-s)/2 + \sqrt{\nu^2-a^2}]^2+a^2}}.
\label{spectrumw}
\ee

One can show that the equation $\omega_{\xi}(E)=0$
has solution  $E=-m$ for $1/2>\gamma\geq 0$ and $\xi\neq 0$.
Thus,  bound fermion (particle) states exist while $\gamma\geq 0 (q\leq q_c)$. For $1/2>\gamma\geq 0$, the energy levels can become  negative and decrease to the lower continuum boundary $-m$  but no fermion states will cross it.

\section{Quasistationary  states and creation of fermions. Massive case}

In the overcritical range $q>q_c (\gamma=i\sigma)$  the left-hand side of (\ref{bounsym1}) is
$$
(\bar f_1f_2-\bar f_2f_1)|_{r=0} = -(2is\sigma/a)(|c_1|^2-|c_2|^2).
$$
Thus, there is one-parameter family of the operators $h_{\theta}$ given by
\bb
h_\theta{:}\left\{\begin{array}{l}
 D_\theta=\left\{\begin{array}{l}
F(r):F(r)\;\mbox{is absolutely continuous in}[0,\infty),
\nonumber \\
F,\check hF\in{\mathfrak L}^2(0,\infty), \\
F(r)=c[e^{i\theta}r^{i\sigma} u_+ +e^{-i\theta}r^{-i\sigma} u_-]+O(r^{1/2}),\\
{r\rightarrow{0}},\quad 0 \le \theta\le \pi,\quad 0\thicksim \pi,
\end{array}\right.\\
h_\theta F=\check hF,
\nonumber
\end{array}\right.
\ee
where $c$ is arbitrary constant. We have taken into account that  $c_2=e^{i\theta}c_1$, $0\leq \theta\leq 2\pi$   is equivalent to $c_1=e^{i\theta}c$, $c_2=e^{-i\theta}c$, $0\leq \theta\leq \pi$ with
replacement $\theta\to 2\pi-2\theta$. For $\gamma=i\sigma$ the doublets $U_\theta(r;E)$ and $V_{\theta}(r;E)$ should be chosen in the form
\begin{align}
&U_\theta(r;E)=e^{i\theta}U_1(r;E)+e^{-i\theta}U_2(r;E),\label{e75}\\
&V_\theta(r;E)=U_\theta(r;E)+
\frac{ia\omega_\theta(E)}{4s\sigma}[e^{i\theta}U_1(r;E)-e^{-i\theta}U_2(r;E)],
\nonumber
\end{align}
where $U_1(r;E)$, $U_2(r;E)$ are determined by (\ref{1e35}) with $\gamma=i\sigma$,
the Wronskian $\omega_{\theta}(E)\equiv {\rm Wr}(U_{\theta},V_{\theta})$ is
\bb
\omega_{\theta}(E)=
-\frac{4is\sigma}{a}\frac{1-\tilde{\omega}(E)e^{2i\theta}}{1+\tilde{\omega}(E)e^{2i\theta}},
\quad \tilde{\omega}(E)=\frac{a}{2si\sigma}\omega(E)
\label{mainm}\ee
and $\omega(E)$ is given by eq. (\ref{wren}) with $\gamma=i\sigma$.
In the range $|E|>m$
the function $\omega_{\theta}(E)$ is
continuous, complex-valued and  not equal to zero for real $E$;
the spectral function $\sigma(E)$ exists and is absolutely continuous and
the energy spectrum  is continuous.
In the range $|E|<m$, let us write $\tilde{\omega}(E)\equiv{e}^{-2i\Omega(E)}$,
 so the function
$\omega_{\Omega}(E)=4s\sigma\tan(\Omega(E)-\theta)/a$ is real and the spectrum is implicitly determined by
\bb
\sigma\ln\frac{2\lambda}{m}+\arg\Gamma(2i\sigma)+\arg\Gamma\left(\frac{1-s}{2}+\frac{aE}{\lambda}
+i\sigma\right)+ \arctan\frac{s\sigma}{\nu+a(m+E)/\lambda}+\theta =k\pi,\quad k=0,\pm 1, \ldots .
\label{spsup}
\ee

One can show that $E=m$ is a spectrum accumulation point and  the number
of discrete energy levels is finite in the interval  $0>E\geq -m$.
For $\mu>0$ the lowest bound state is the state with $s=-1$.
For $0<\sigma\ll 1$ eq. (\ref{spsup}) has real solution $E=-m$ for $k=0$  and $\theta=\pi/2-\sigma\ln 2a_c-\arctan(s\sigma_c)/\nu+\sigma_c{\cal C}$, where
${\cal C}=-\psi(1)=0.57721$ is the Euler constant and $\psi(z)$ is the logarithmic derivative of Gamma function  \cite{GR}.

As we introduce a small change in $\sigma$ such that $\sigma>\sigma_c$
there is a sudden change in spectrum: there are no solutions of Eq. (\ref{spsup}) for real $E$. Therefore, one of the bound state poles disappears from
the physical sheet: for $E<-m$ only the continuous spectrum exists, but below
${\rm Re}\lambda>0, {\rm Im}\lambda>0$ there is a second (unphysical) sheet ${\rm Re}\lambda<0, {\rm Im} \lambda<0$ on which the virtual bound state pole resides at $\sigma>\sigma_c$. The key difference of the case $\sigma>\sigma_c$ is that virtual bound states have ``complex energies'' $E=|E|e^{i\tau}$, which are determined by complex equation $\omega_{\theta}=0$:
\bb
\frac{\Gamma(2i\sigma)}{\Gamma(-2i\sigma)}\frac{\Gamma(-i\sigma+(1-s)/2-iaE/p)}
{\Gamma(i\sigma+(1-s)/2-iaE/p)}\frac{(-2ip)^{-2i\sigma}}{m^{-2i\sigma}} \frac{\nu+i[a(E+m)/p+s\sigma]}{\nu+i[a(E+m)/p-s\sigma]}={e}^{-2i\theta},
\nonumber \\
 p=\sqrt{E^2-m^2}.\phantom{mmmmmmmmmmmmmmm}
\label{comeq}\ee
 For ${\rm Re}E=-(m+\epsilon), \epsilon\to +0, \quad 1 \gg \sigma>0$, one obtains
$[1-2\sigma{\rm Im} \psi(-iz)]e^{-\pi\sigma+2\sigma\alpha}=1$ and
$\alpha \approx \pi+(\pi/2)e^{-\sqrt{2m\pi a^2/\epsilon}}$,
as well as
\bb
 \arg\Gamma(2i\sigma)-\sigma{\rm Re}\psi(-iz)-(\sigma/2)\ln(8\epsilon/m)+(1/2)\arctan[s\sigma(1-a^2\epsilon/2m\nu^2)/|\nu|]=-\theta+\pi n,
\label{bound-men}
\ee
where $z=\sqrt{ma^2/2\epsilon}$. The first quasistationary state emerges when the ground bound state with $s=-1, l=0$ "dives"  into the lower continuum. There appears the pole on the unphysical sheet,
counted now as a ``positron'' state.

Putting ${\rm Re}\psi(-iz)\approx -{\cal C}+\ln z,\quad n=0$,
 we obtain:
\bb
 \frac{s\sigma a^2\epsilon}{2m|\nu|^3}=-[{\cal C}+\ln(2a_c)+ \sigma/(2\sigma_c)-s/|\nu|](\sigma-\sigma_c).
\label{bounden0}
\ee
The physical picture can be seen as follows.
When $\sigma>\sigma_c$, the lowest energy level dives into the
negative energy continuum and becomes a resonance. It is spread
out over an energy range of the order
$\Gamma_g \sim me^{-\sqrt{2m\pi a^2/\epsilon}}$ and strongly distort
around the impurity. The width $\Gamma_g$ is the doubled probability of the creation of the electron–positron pair by the Coulomb potential in the presence of AB potential.  It is exponentially small in this case.
The additional distortion of the negative
energy continuum (due to the diving bound state)  leads to a negative charge
density due to the ``real vacuum polarization'', since its origin is not a fluctuating pair or the K-shell bound electron state (for $a<a_c$), but the structured vacuum of supercritical QED \cite{grrein}.
The diving point for the energy level defines
and depends upon the  parameter $\theta$.

The resonance is not
usual bound level diluted inside a continuum, where
its lifetime essentially disappears. The
overcritical level remains sharply defined with diverging
lifetime $\tau\sim e^{\sqrt{2m\pi a^2/\epsilon}}/m$.
The resonance is practically a bound state.
This diving of bound levels entails a complete restructuring
of the vacuum. If the emergent level was empty, an electron--positron
pair will be created: the electron from the Dirac sea occupies
this  level  and shields the charge of the source,
while the  positron (hole) escapes to infinity \cite{zelpop,grrein}.
As a result, when $\sigma>\sigma_c$ the QED vacuum  acquires the charge
$e$ \cite{khho}, thus leading to the concept of a charged
vacuum in overcritical fields due to the real vacuum polarization \cite{zelpop,grrein}.
An essential detail is that the vacuum charge spatial
distribution is similar to the modulus squared of the fermion wave function
in the lowest bound state. However, the modulus squared of the fermion wave function is the probability of  finding the charge (equal to $e$) at a given spatial point $r$
while the vacuum charge density characterizes the spatial distribution of the real electric charge appearing in the vacuum. The spatial distribution of the real vacuum charge is
at $r\to 0$
$$
e|\Psi(r)|^2\sim em[2(\ln mr-\xi)^2-2s(\ln mr-\xi)/a +1/a^2]
$$
and at $r\to \infty$
$$
e|\Psi(r)|^2\sim ee^{-2\sqrt{r/l}}/r, \quad l=1/\sqrt{2m\epsilon},
$$
where $\epsilon$ depends upon $\xi, \gamma, \mu, a$.
In 2+1 dimensions, the QED vacuum can also acquire a magnetic moment equal to the
spin magnetic moment of the electron.
 Other levels will sequentially follow at higher $\sigma_c$.

\section{Quasistationary  states and creation of massless fermions}

The massless fermions do not have spin degree of freedom in 2+1 dimensions \cite{jacknai}.
Nevertheless, the Dirac Hamiltonians in the  AB potential  for charged massless fermions in 2+1 dimensions keep the introduced spin parameter.
So, all obtained solutions (doublets) are valid for the case $m=0$
in the corresponding charge ranges if we put: $m=0$,
$x = -2i|E|r$, $a^s=\gamma_s+(1-s)/2-ie'a$,  $e'=E/|E|$,
$m_s^{\pm}=(s\gamma-ie'a)/\nu$.
The main Wronskian $\omega(E)={\rm Wr}(U_1,V_1)$  at $m=0$ takes the form
\begin{align}
\omega_0(E)=&\frac{\Gamma(2\gamma)\Gamma\left(-\gamma+(1-s)/2-ia\right)}{\Gamma(-2\gamma)
\Gamma\left(\gamma+(1-s)/2-ia\right)}(-2iE)^{-2\gamma}\frac{\nu+ia+s\gamma}{\nu+ia-s\gamma}\frac{2s\gamma}{a}\equiv \frac{\tilde{\omega}(E)}{\Gamma(-2\gamma)}.
\label{wrm0}
\end{align}

For $0<\gamma<1/2$ now the energy spectrum is determined by $\omega_\xi^0(E)={\rm Wr}(U_\xi,V_\xi)=\omega_0(E)+2s\gamma\xi/a$ with $\omega_0(E)$  (\ref{wrm0}).
It can be verified that in the range $|E|>0$
the functions $\omega_0(E)$  and $\omega_{\xi}^0(E)$ are
continuous, complex-valued and  not equal to zero for real $E$;
the spectral function $\sigma(E)$ exists and is absolutely continuous.
The energy spectrum in the range $|E|>0$ is continuous and the quantum system under discussion does not have bound states. Nevertheless, $E(a,\nu,s,\xi)$ determined by equation ${\rm Re}\omega_{\xi}^0(E)=0$
\bb
E=\frac{e'}2\left[\frac{\Gamma(1+2\gamma)|\Gamma(-\gamma-ia)|}{|\xi|\Gamma(1-2\gamma)|\Gamma(\gamma-ia)|}
\sqrt\frac{\nu+s\gamma}{\nu-s\gamma}\right]^{1/2\gamma}
\label{boundri10sol}\ee
 may characterize some kind of accumulation points of fermion states
and the corresponding values $a, \nu, s, \xi$ for these points must satisfy equation (${\rm Im}\omega_{\xi}^0(E)=0$)
\bb
 \pi\left(e'\gamma-\frac12\right)-\frac{3+s}{4}\arctan\frac{4a\gamma}{4\gamma^2-(1+\nu^2)(1-s)}+
 \nonumber \\
 +\sum\limits_{n=1}^{\infty}\arctan\frac{8a\gamma}{(2n+1-s)^2+4(a^2-\gamma^2)}=(p-1)\frac{\pi}{2},
\label{boundri20}\ee
where $p=\xi/|\xi|=\pm 1$, $p=1 (-1)$ for $\infty>\xi\geq 0 (0\geq \xi>-\infty)$.

We shall put $\mu>0$. The case  $\mu<0$  can be discussed similarly with the signs of $l$ and $s$ flipped. The  range near $|E|=0$ is of interest.
For  $\gamma\to 1/2$
\bb
E=e'\frac{1-2\gamma}{2|\xi|}\frac{|\Gamma(-1/2-ia)|}{|\Gamma(1/2-ia)|}
\sqrt\frac{\nu+s/2}{\nu-s/2},
\label{newb}
\ee
hence $|E|=0$  and eq. (\ref{boundri20}) is satisfied
by $\gamma=1/2$ for $e'=1, p=1 (\pi\geq \theta\geq 0)$ and for $e'=-1, p=-1 (2\pi\geq \theta\geq \pi)$ only when  $a^2=\nu^2-1/4$, i.e. at $\mu=0$ only for $a=0$ (compare with claim in \cite{vpnn}). There is the particle-hole symmetry in free particle case ($a, \mu=0$).

For $\gamma\to 0$, $|E|$ tends to $0$  as
$2E\approx e'(1/|\xi|)^{1/2\gamma}$ and (\ref{boundri20})
 is satisfied by $e'=\pm 1$, $\gamma=0$ only for $p=-1(0\geq\xi>-\infty,\;2\pi>\theta\geq3/2\pi)$.
This means that the fermion states heap up close to the point $E=0$ for $E>0$ and,
conversely, for $E<0$ only when $|\xi|>1$ (see also \cite{vpnn})
but no fermion states will cross it  as well as no virtual bound states exist while $q<q_c$.
For $\gamma=i\sigma$ the point $E=0$ is the branch point of the  Wronskians
in the complex plane of $E$; the quasistationary states situate
on the unphysical sheet. For  $m=0$ the main Wronskian has the form (\ref{mainm})
in which $\omega(E)$ is given by (\ref{wrm0}) with $\gamma=i\sigma$.
One can verify again that  $\omega_{\theta}^0(E)$ are
continuous, complex-valued and is not equal to zero  for real $E$,
so no bound  states exist. Physically, this is because
there is no natural length scale in the problem to characterize
bound states. Nevertheless, the  virtual bound (resonant)  states can
emerge when $q>q_c$;  their complex ``energies'' $E=|E|e^{i\alpha}$  are determined by:
\bb
\frac{|\Gamma((1-s)/2-i(a+\sigma))|}{|\Gamma((1-s)/2-i(a-\sigma))|}
\sqrt\frac{a+s\sigma}{a-s\sigma}e^{-\pi\sigma+2\sigma\alpha}=1
\label{boundsig10}\ee
and the equation for the energy spectrum
\begin{align}
 &2\sigma\ln(|E|/E_0)= 2\theta -\pi\left(1+2k\right) -2\sigma{\cal C}+\arctan\frac{s\sigma}{\nu}+ \sum\limits_{n=1}^{\infty}\left(\frac{2\sigma}{n}-
 2\arctan\frac{2\sigma}{n}+\arctan\frac{2\sigma n}{n^2+\nu^2}\right).
\label{boundover}
\end{align}
where a positive constant $E_0$ gives an energy scale and $\pi\geq \theta\geq 0$.
It should be emphasized that now $e'=1$ ($e'=-1$) also corresponds to the physical sheet (the unphysical sheet). Increasing $a$ ($\sigma$) will increase  $k$ and decrease the energy.
 This has to do with the fact that, in reality, the Dirac
point is an accumulation point of infinitely many resonances \cite{vpnn}.

For  $\sigma\ll 1$,  eq. (\ref{boundsig10}) has approximate  solution
$\alpha \approx-(1+s)/4a+\mathrm{Im}\psi(ia)+\pi/2$ and
  $\alpha \approx [1+\coth(\pi/2)]\pi/2\approx (1+0.04)\pi$ for $a=1/2, s=1$.
Eqs. (\ref{boundsig10}) and (\ref{boundover})
are approximately satisfied near $|E|=0$ only
for hole region $E<0$. Indeed, for $a>\nu$, $\sigma>0$
eq. (\ref{boundsig10}) is satisfied  only at  $e'=-1, \tau>\pi$ for which
  the right hand side of (\ref{boundover})
is negative.  Then, for $\sigma\ll 1$ the energy spectrum is
\begin{align}
 &E_{k,\theta,s}= E_0 \cos(\alpha)
 \exp\left[-\pi(1+2k)/2\sigma+\theta/\sigma-({\cal C}+(1-s)/2
 +\pi^2/6-(\pi\coth\pi a)/2a)\right].
\label{energyres}
\end{align}
 These energies have an essential singular point at $\sigma=0$  \cite{ashkl,kupgc,ggg}.
The infinite number of quasistationary levels is
related to the long-range character of the Coulomb potential \cite{ggg,vpnn,ashkl}.
These quasi-localized resonances have negative energies, thus they are situated in the hole sector.
The resonances  are directly associated with the positron production in  the QED \cite{nkpnp}.

The imaginary part of $E_{k,\theta,s}$ defines the width of
virtual resonant level $\Gamma_{k,\theta,s}$ or the inverse lifetime (decay rate) of particle resonance. For $\sigma\ll 1$
this width   $\sim |E_{k,\theta,s}|$ is very small, hence,
the resonances are practically stationary states.

The spectrum in the case of charged massless fermions is continuous
everywhere, and so there is no restructuring of negative
energy (hole) continuum  in overcritical fields
due to the real vacuum polarization as described
for the massive case.  The physical picture can be seen as follows.
If the emergent virtual level was empty, a
quasiparticle pair will be created: the fermion (particle) of the filled valence band
occupies this level and shields the center,
while the emergent (in the valence band) hole is escaped to infinity.
Now the quantity $\Gamma_{k,\theta,s}$ is the doubled probability of
the creation of the quasiparticle pair by the Coulomb potential
in the presence of AB potential.

The physically meaningful quantity is the number
of pairs created per unit area of graphene per unit time.
So, the creation of massless charged fermions can be studied by means
of the local density of states (LDOS) in the hole continuum.
The LDOS per unit area is determined as a function
of energy  and distance from origin by \cite{vpnn}
\begin{align}
 N(E,r)&=\sum\limits_{l}|\Psi(t,{\bf r})|^2=\sum\limits_{l}n_l(E,r), \quad
 n_l(E,r)=\frac{|f_1(r,E,l)|^2+|f_2(r,E,l)|^2}{2|A_l(E)|^2\pi r},
\label{dens}
\end{align}
where $f_1(r,E,l)/A_l(E)$ and $f_2(r,E,l)/A_l(E)$ are the doublets  normalized
(on the half-line with measure $dr$) by imposing orthogonality on the energy scale and
$A_l(E)$ is the normalization constant.

The LDOS is: 1. $N_{reg}(E,r)=\sum\limits_{l}n_l(E,r)$ with  $n_l(E, r)$ constructed by regular solutions of (\ref{1e35}) and
with the sum  taken over $l$ of  $\sqrt{(l+\mu+s/2)^2-a^2} \geq 1/2$ for $\gamma\geq 1/2$.
For $a=0$, $\mu=0$ the free density of states  is recovered from $N_{reg}(E,r)$  to be  $N(E,r)=|E|/2\pi$;   2. $N_{\xi}(E,r)= \sum\limits_{l}n_l(\xi, E,r)$ with the sum taken over $l$ of $1/2>\sqrt{(l+\mu+s/2)^2-a^2}>0$ for $1/2>\gamma>0$;
3. $N_{\theta}(E,r)=\sum\limits_{l} n_l(\theta, E,r)$
with  $n_l(\theta, E, r)$  constructed by (\ref{e75}) and the sum taken over $l$ of  $a^2>(l+\mu+s/2)^2$ for the overcritical range  $\gamma=i\sigma$, $0\geq \theta\geq \pi$.
 The total LDOS  is $N(E,r)=N_{reg}(E,r)+N_{\xi}(E,r)+N_{\theta}(E,r)$.

\begin{figure}[h!]
\centering
{\includegraphics[width=4.2cm]{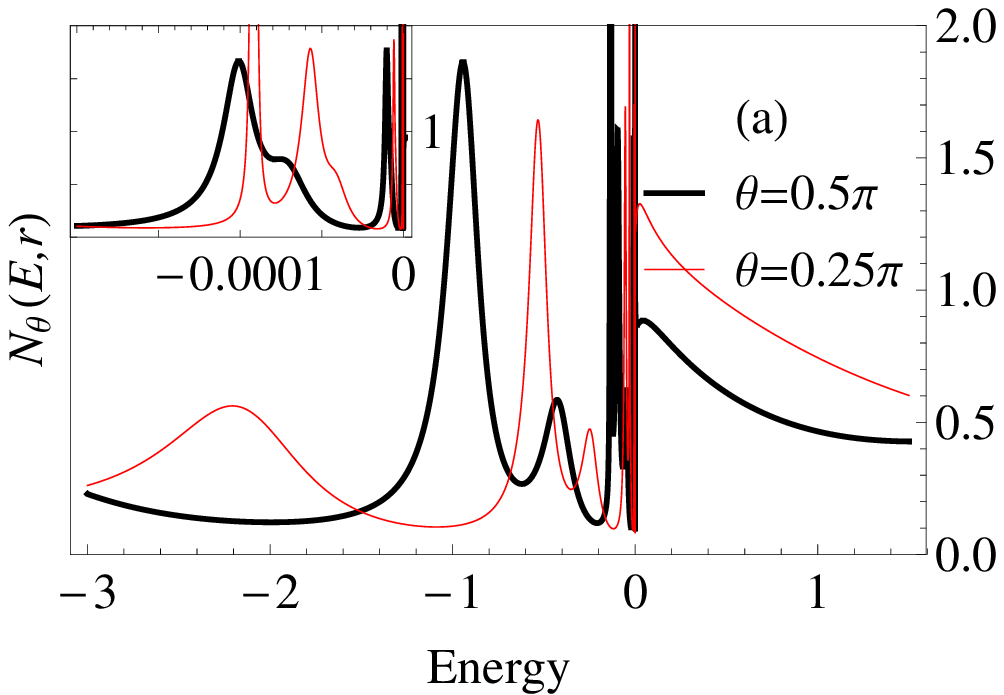}}
{\includegraphics[width=4.2cm]{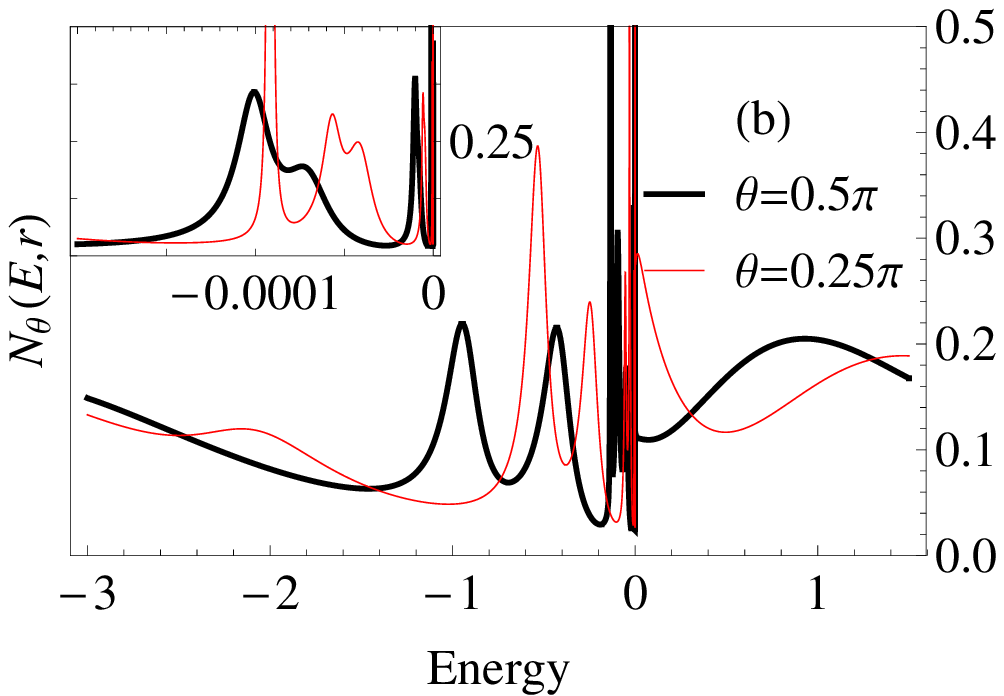}}
\caption{LDOS $N_\theta(E,r)$ with $l=-2,-1,0$ for $a=1.5,\mu=0.1, s=1$
($\sigma\approx 0.539, 1.446, 1.375$) and $r=0.3\;({\rm a}),
r=1\;({\rm b})$; the insets are magnifications for $E\approx 0$.}
\label{fig.1}
\end{figure}

\begin{figure}[h!]
\centering
{\includegraphics[width=4.2cm]{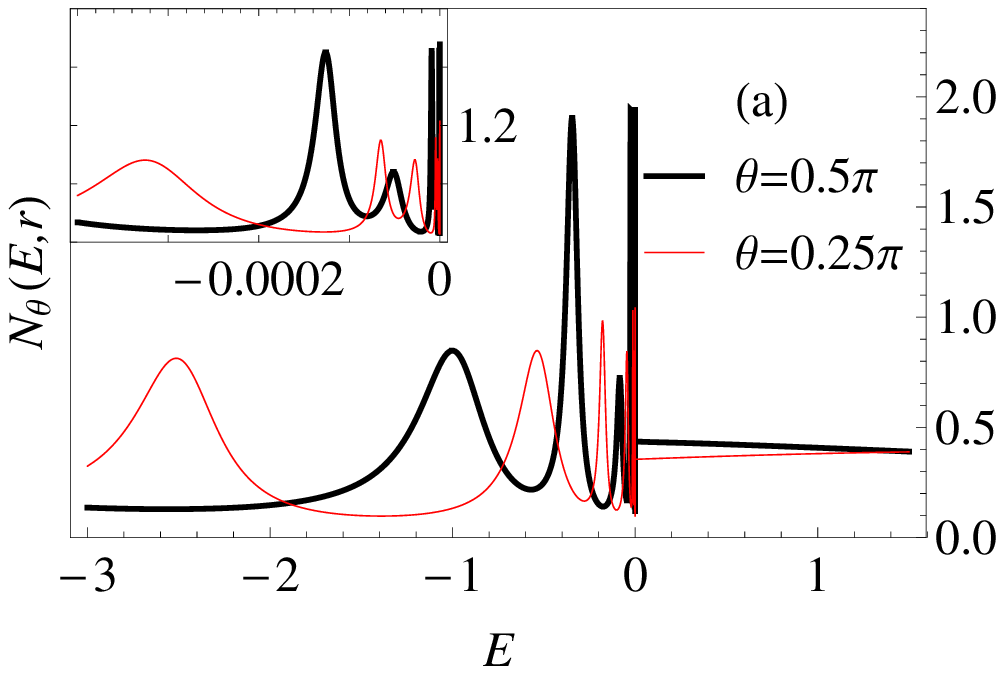}}
{\includegraphics[width=4.2cm]{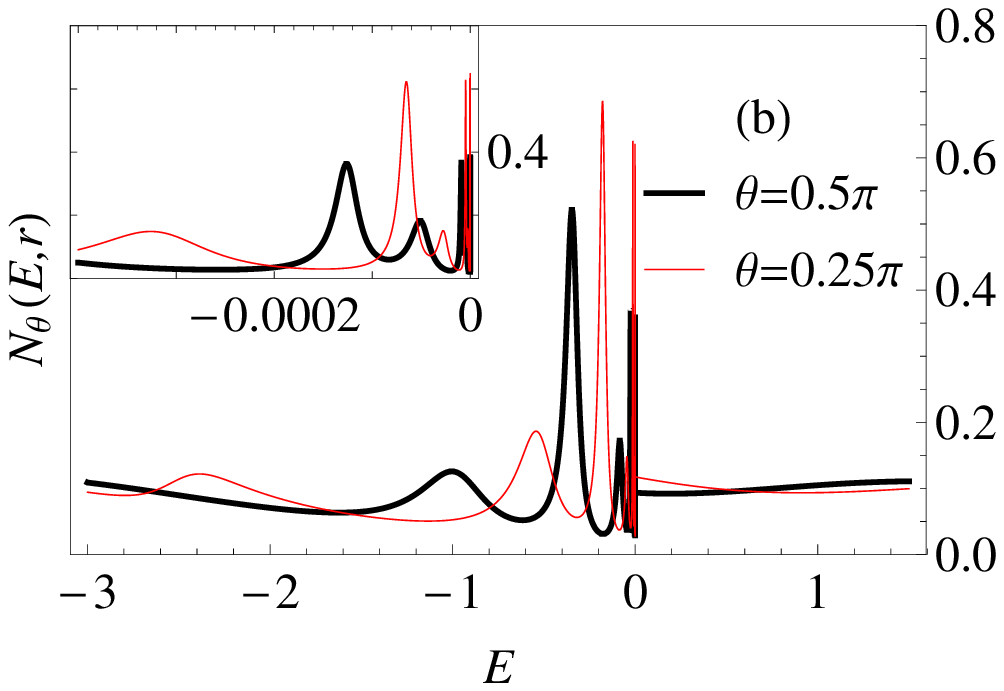}}
\caption{$N_\theta(E,r)$ with $l=0 (\sigma\approx1.272), 1(\sigma\approx1.191), a=4/3, \mu=0.1, s=-1$ and $r=0.3$ (a), $r=1$ (b); the insets are magnifications for $E\approx 0$.} \label{fig.3-s}
\end{figure}

The LDOS  exhibits  resonances of the width  $\sim |E_{k,\theta,s}|$ at the negative
energies (\ref{energyres}), which decay away
from the impurity (Figs.~\ref{fig.1} for $s=1$ and \ref{fig.3-s} for $s=-1$); strong resonances
 signal the presence of quasistationary  states, i.e. the creation of charged fermions.
 It should be commented that, according to (\ref{grel}), the families of the curves for the LDOS
with another sign $s$ are qualitatively like
to the ones given in Figs. 1 and 2 at the same values of $a, \mu, \xi, \theta$ except to the shift $\pm l\to \pm l+1, s\to -s$.

Increasing the effective charge will cause energy quasiparticles
to decrease and their number to increase.

\begin{figure}[h!]
\centering
{\includegraphics[width=4.2cm]{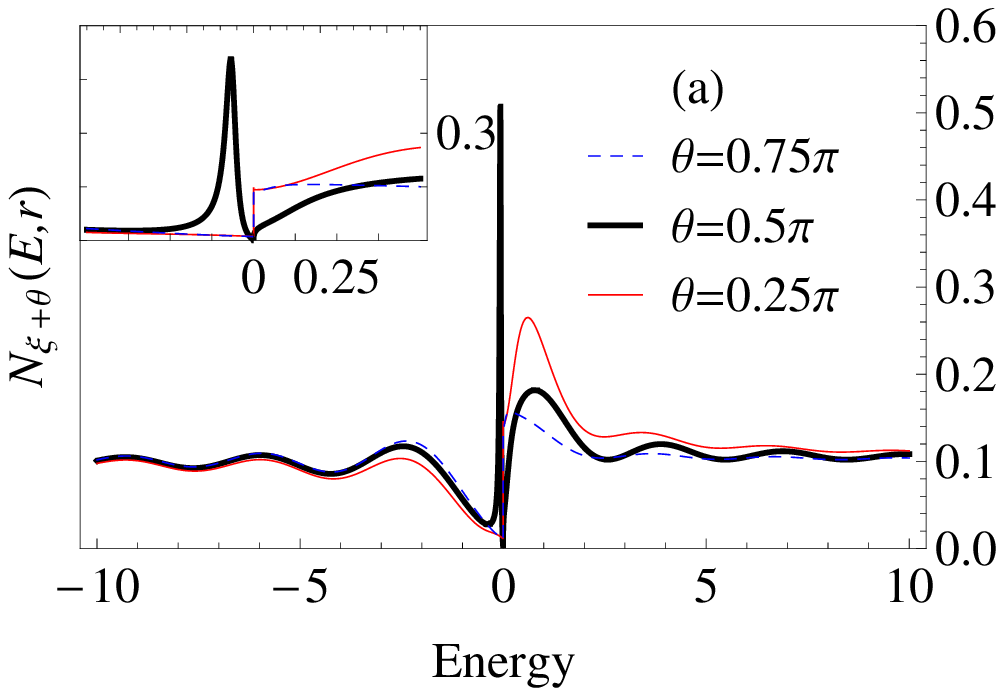}}
{\includegraphics[width=4.2cm]{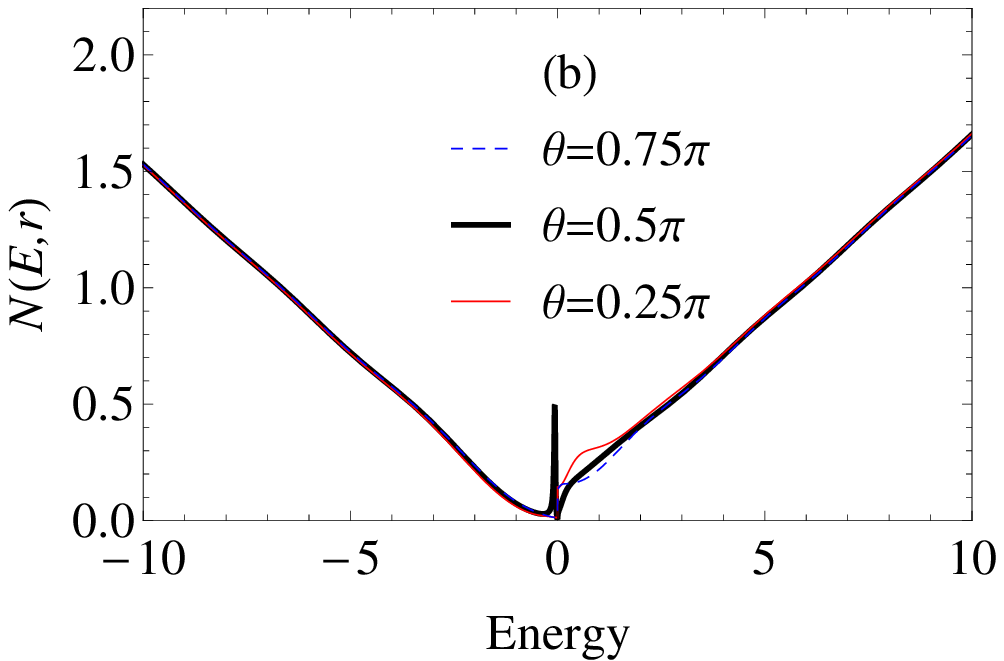}}
\caption{$({\rm a})$  $N_{\xi+\theta}(E,r)=N_\xi(E,r)+N_\theta(E,r)$ for $l=0$ ($\gamma\approx0.4472$) and $l=-1$ ($\sigma\approx0.0028$); the inset is a magnification for $E\approx0$. $({\rm b})$ $N(E,r)=N_{reg}(E,r)+N_\xi(E,r)+N_\theta(E,r)$. On all panels $a=0.40001, \mu=0.1, s=1, r=1$.}
\label{fig.4}
\end{figure}
Figure \ref{fig.4} shows that when $\sigma\to 0$ there exists a single resonance (with $k=0$ at $\theta =\pi/2$, and only for $s=1$), which is in good accord with (\ref{energyres}).

\section{Summary}

We have investigated the creation of charged fermions from the vacuum
by a Coulomb field in the presence of an Aharonov--Bohm
potential in 2+1 dimensions  taking into account the physical effects
 due to the interaction between the electron spin
magnetic moment and the AB magnetic field.

For the massive case,  the lowest energy level dives into the
Dirac sea and turn into a resonance with ``complex energy'' in
the overcritical range; there appears the pole on the unphysical sheet
counted as a ``positron'' state. The diving point as well as effective critical charge  define
and depend upon the extension parameter $\xi$.
Other virtual bound states will sequentially emerge
at higher effective critical charges.
The vacuum of the quantum electrodynamics becomes unstable,
which results in positron creation; it  is
reconstructing: a new  state with the energy $E < -m$ emerges and is spread out over
an energy range of the order $\Gamma_g \sim me^{-\sqrt{2m\pi a^2/\epsilon}}$.
   The critical charge, respectively,
decreases (increases) at fixed $a$ in the presence of magnetic  flux with $\mu>0$  for $s = -1$ ($s = 1$).   This means that the vacuum of the quantum electrodynamics
in 2+1 dimensions in Coulomb and AB potentials   with $\mu>0$
becomes less stable with respect to the creation of positrons with the spin $s_p = 1$ and more stable with respect to the creation of positrons with the spin $s_p = -1$.

The creation of massless charged quasiparticles in Coulomb and AB fields
in graphene  differs with the case of massive particles as follows:
(i) there is no restructuring of the hole (lower) continuum, (ii) in some range
of extension parameter there  reveals an infinite number of quasistationary states
at $\sigma>\sigma_c$ in the lower continuum, (iii) when the mass $m=0$ there
is no natural length scale  to characterize such quasistationary states.

It will be noted that at the moment graphene single crystals with characteristics (such as dimensions, electron mobility or concentration of impurities), which is favorable enough for observation of the effect of quasiparticles creation,  are obtained \cite{xlcman}.

\vspace{1cm}

{\bf Acknowledgments}. The author is grateful to K. E. Lee for the help with the numerical calculations.

\end{document}